\documentclass[11pt]{article}
\usepackage{graphicx}
\usepackage{latexsym,graphicx}
\usepackage{float}
\usepackage{amsmath}
\usepackage{amsfonts}
\usepackage{amssymb}
\usepackage{epstopdf}
\usepackage{color}
\DeclareGraphicsExtensions{.eps}
\usepackage[left=2.5cm,top=2.5cm,right=2.5cm,bottom=2.5cm]{geometry}

\catcode `\@=11
\catcode `\@=12
\begin{document}
	\begin{center}
	\large{{\bf The Reconstruction of Constant Jerk Parameter with $f(R,T)$ Gravity in Bianchi-I spacetime}} \\
	\vspace{5mm}
	\normalsize{Anirudh Pradhan$^1$, Gopikant Goswami$^2$, Syamala Krishnannair$^{3}$}\\
	\vspace{5mm}
	\normalsize{$^{1}$Centre for Cosmology, Astrophysics and Space Science (CCASS), GLA University, Mathura-281 406, Uttar Pradesh, India}\\
	\vspace{5mm}
	\normalsize{$^{2}$Department of Mathematics, Netaji Subhas University of Technology, Delhi, India}\\
	\vspace{5mm}
	\normalsize{$^{3}$Department of Mathematical Sciences, University of Zululand Private Bag X1001 Kwa-Dlangezwa 3886 South Africa}\\
		\vspace{2mm}
	$^1$E-mail: pradhan.anirudh@gmail.com \\
	\vspace{2mm}
	$^2$E-mail: gk.goswami9@gmail.com \\
	\vspace{2mm}
	$^{3}$E-mail:krishnannairs@unizulu.ac.za\\
	\end{center}
	\begin{abstract}
		 We have developed a Bianchi I cosmological model of the universe in $f(R,T)$ gravity theory which fit good with the present day scenario of accelerating universe.  The model displays transition from deceleration in the past to the acceleration at the present. As in the  $\Lambda$CDM model, we have defined the three energy parameters  $\Omega_m$, $\Omega_{\mu}$ and $\Omega_{\sigma}$ such that $\Omega_m$ + $\Omega_{\mu}$ +  $\Omega_{\sigma}$ = 1. The parameter $\Omega_m$ is the matter energy density (baryons + dark matter), $\Omega_{\mu}$ is the energy density associated with the Ricci scalar $R$ and the  trace $T$ of the energy momentum tensor and $\Omega_{\sigma}$ is the energy density associated with the anisotropy of the universe. We shall call $\Omega_{\mu}$  dominant over the other two due to its higher value.  We find that the $\Omega_{\mu}$  and the other two  in the ratio 3:1.   46 Hubble OHD data set is used to estimate present values of Hubble $H_0$, deceleration $q_0$ and jerk $j$ parameters. 1$\sigma$, 2$\sigma$ and 3$\sigma$  contour  region plots for the estimated values of parameters  are presented. 580 SNIa supernova distance modulus data set and  66 pantheon SNIa data which include high red shift data in the range $0\leq z\leq 2.36$ have been used to draw error bar plots and likelihood probability curves for distance modulus and  apparent magnitude of SNIa supernova's.  We have  calculated the pressures and densities associated with the two matter densities, viz., $p_{\mu}$, $\rho_{\mu}$, $p_m$ and $\rho_m$, respectively. The present age of the universe as per our model is also evaluated and it is found at par with the present observed values.

\end{abstract}


 
\section{Introduction:}

A new era in cosmology had begun almost two and half decades earlier when the concept of accelerating  universe and a bizarre type of hidden energy which is given the name dark energy(DE), surfaced in the literature \cite{1} $-$ \cite{9}. It is said that DE at present is dominating the universe and it is responsible for creating the acceleration in the universe due to its property of having negative pressure. Many cosmological models have been proposed assuming DE as a perfect fluid with negative pressure \cite{10} $-$ \cite{14}. As cosmological constant($\Lambda$) has property of producing negative pressure, $\Lambda$CDM model was resurrected and found best fit on observational basis but it lacks in some theoretical grounds \cite{15} $-$ \cite{20}. As an alternative, quintessence and phantom models \cite{21} $-$ \cite{25} were developed in which a tracker scalar field $\phi$ producing negative energy was proposed as a dark energy. Some more  dynamical dark energy models with a variable equation of state parameter were developed.  Along with these some  holographic dark energy (HDE) models\cite{26} $-$ \cite{31} have been developed, but they do have limitations and face non conservation of energy problem \cite{32} $-$ \cite{35}.\\

In the year 2011, $f(R,T)$ gravity theory proposed by Harko et al. \cite{36} gave a new direction to the researchers that  the modification of Einstein Hilbert action by replacing Ricci scalar $R$ and placing an arbitrary function of $R$ and trace $T$ of energy momentum tensor may develop  effective cosmological constant type negative pressure and as a result may produce acceleration in the universe. Off late the authors have developed accelerating FLRW cosmological model in $f(R,T)$ gravity. Some important applications and reviews of $f(R,T)$ gravty can be found in \cite{37} $-$ \cite{44}. Earlier to it, Nojiri and Odintsov \cite{45} have replaced Ricci scalar $R$ in the action of  General relativity  by an arbitrary function $f(R)$  of $R$  and investigated a theoretical cosmological model. These models successfully described the late time acceleration of the Universe \cite{46}, \cite{47}. In some more   Refs. \cite{48} $-$ \cite{53}, we find viable cosmological models in the $f(R)$ theory of gravity fit best on the solar system test and the galactic dynamic of massive test particle without inclusion of dark matter. Some other modified theories of gravity such as $f(G)$ \cite{54}, $f(R,G)$ \cite{55} and $f(T,B)$ \cite{56} theories also surfaced in the literature.
 Wilkinson Microwave Anisotropic Probe (WMAP) and its findings \cite{57}, \cite{58} indicates that our universe is not totally isotropic, it has small fluctuation in density along directions. This has created interest in Bianchi type anisotropic models.  Akarsu et al. \cite{59} have developed Bianchi type I universe model with observational constraints. In references \cite{60}-\cite{65}, we will see some interesting accelerating Bianchi-I cosmological models  with observational data sets fitted with them. In order to realize bounce cosmology, Bamba et al. \cite{65a} constructed F(R) gravity models with experimental and power-law forms of the scale factor. Additionally, an f(R) bigravity model realizing the bouncing behavior is reconstructed in \cite{65a} for an experimental form of scale factor. By taking into account the Klein-Gordon equation in the f(R) theory of gravity and using the various values of EoS parameters, Malik and Shamir \cite{65b} have looked at some numerical solutions for FRW space-time. Shamir and Meer \cite{65c} described a comprehensive study of relativistic structure in the context of recently proposed  $\Re + \alpha \mathcal{A}$ gravity, where $\Re$ is the Ricci scalar, and $\mathcal{A}$ is the anti-curvature scalar. In this study they \cite{65c} examined a new classification of embedded class-I solutions of compact stars. In the context of f(R) gravity, the LRS Bianchi type I universe is investigated, and reveal three solutions with corresponding Killing symmetries in \cite{65d} and \cite{65e}. Shamir et al. \cite{65f} studied the relativistic construction of compact astronomical solutions in f(R) gravity and found that the Bardeen model of black holes describes  physically realistic stellar structures.	Recently, Shamir \cite{65g} studied the bouncing cosmology in f(G,T) gravity with lagarithmic trace term and provided bouncing solutions with the chosen EoS parameter. Anisotropic universe in f(G,T) gravity is also discussed in \cite{65h}. \\
 
 In this research article, we attempt to develop a cosmological model of the universe which is spatially homogeneous and anisotropic. For this we consider a Bianchi type I space time in which scale parameter $a$ is function of time but its rate of expansion is different at different directions. The objective is there to work with a model which fits suitably with the presently available data sets. For this purpose, we use $f(R,T)$ gravity and its field equations in the background of Biachi I space time. We consider the simplest form of $f(R,T)$ as  $f(R,T) = R + 2 \lambda T$, where $\lambda$ is an arbitrary constant to comply our objective. We are using three types of data sets: latest compilation of $46$ Hubble  data set,  SNe Ia $580$ data sets of distance modulus and   66 Pantheon data set of apparent magnitude which comprised of 40 SN Ia  bined and 26 high redshift data's in the  range $0.014 \leq z \leq 2.26 $. There is a jerk parameter $j$ related with third order of derivatives of scale factor. It is used in examining instability of a cosmological model and in statefinder diagnostic. In $\Lambda$CDM model $j=1$ and in other models where jerk is variables, it varies very slowly. For instance, in \cite{66}, jerk vary from $0.99 -1.04$, so we model an Universe with constant jerk parameter in $f(R,T)$ gravity. We have developed a universe model with constant jerk parameter $j$. Our model is at present accelerating and it shows transition from deceleration in the past to acceleration at present. We have statistically estimated jerk $j$, the present values of deceleration $q_0$ and the Hubble parameter $H_0$ with the help of OHD data set. The estimation of $H_0$ is also carried out by the other two data sets mentioned as above. It is found that our model fits well  latest observational findings \cite{67}. We have discussed some of the physical aspects of the model in particular age of the universe.\\

 The paper is presented in the following section wise form. In section $2$, the $f(R,T)$ field equations have been presented and are solved for spatially homogeneous and anisotropic Bianchi I space time by taking  energy momentum tensor as that of perfect fluid which occupies the universe. We arrive at the two equations of motion of a fluid particle as they contains second order deceleration parameter with the pressure and the Hubble parameter (rate of expansion) with density. We do have one extra term of anisotropy in them. In this section, we expressed density, pressure and equation of state parameter $\omega$ as function of the Hubble and decelerating parameter so once we find the Hubble and decelerating parameter, we may get them.  In section $3$, the deceleration and the Hubble parameter were obtained by taking the jerk parameter constant. It is interesting that decelerating parameter shows transition from negative to positive over red shift which means that in the past universe was decelerating and at present it is accelerating. In section $4$, 46 Hubble OHD data set is used to estimate present values of Hubble $H_0$, deceleration $q_0$ and jerk $j$ parameters. 1$\sigma$, 2$\sigma$ and 3$\sigma$  contour  region and likely hood plots for the estimated values of parameters  are presented. In section $5$, 580 SNIa supernova distance modulus data set and  66 pantheon SNIa data which include high red shift data in the range $0\leq z\leq 2.36$ have been used to draw error bar plots and likelihood probability curves for distance modulus and  apparent magnitude of SNIa supernova's. In section $6$, we have evaluated $\mu$ on the basis of energy parameters $\Omega_{\text{m0}}$, $\Omega_{\mu}$  and  $\Omega_{\sigma}$.  We have obtained $\mu = 0.663073$.  In section $7$,  density, pressure and the equation of state parameter $\omega$ were solved and we have presented graphs for them. In section $8$, we have defined and discuss effective density and pressure and presented graphs. The effective pressure is found negative at present, so it describes acceleration  in the universe. In subsection $8.1$ we have found age of the universe and transition time at which the universe starts accelerating from being decelerating.
 In the last Section $9$, conclusions are provided.
 
 
\section{f(R,T) gravity field equations for FLRW flat spacetime:}

We begin with 
Einstein-Hilbert action:
\begin{equation}{\label{1}}
	S= \int\left(\frac{1}{16\pi G}( R+2\lambda) + L_m \right)\sqrt{-g} dx^4,
\end{equation}
and
Einstein GR field equations:
\begin{equation}\label{2}
	R_{ij}-\frac{1}{2} R g_{ij}+ \Lambda g_{ij} = \frac{8\pi G}{c^4}T_{ij}.
\end{equation}
In $f(R,T)$ gravity action,  Ricci scalar $R$ is replaced by arbitrary function $f(R,T)$ of $R$ an $T$, so it is defined as:
\begin{equation}{\label{3}}
	S= \int\Big(\frac{1}{16\pi G} f(R,T)+L_m\Big)\sqrt{-g} dx^4,
\end{equation}
and on varying the action with respect to metric tensor $g_{ij}$, $f(R,T)$ gravity field equations are obtained as:
\begin{equation} {\label{4}}
	R_{ij}-\frac{1}{2} R g_{ij} = \frac{8 \pi G T_{ij}}{f^R(R,T)}+ \frac{1}{f^R (R,T)} \bigg(\frac{1}{2} g_{ij} (f(R,T)-R f^R (R,T)) - (g_{ij} \Box - \nabla_i \nabla_j) f^R (R,T) + f^T (R,T) (T_{ij} +p g_{ij}) \bigg),
\end{equation}
where the energy momentum tensor $T_{ij}$ is related to the matter Lagrangian $L_m$ via the following equation
\begin{equation}{\label{5}}
	T_{ij}= -\frac{2}{\sqrt{-g}}\frac{\delta(\sqrt{-g} L_m)}{\delta g^{ij}},
\end{equation}
and  $ f^R $ and $ f^T $ are the derivative of $ f(R,T) $ with respect to $ R $ and $ T $ respectively. Three particular functional  forms for arbitrary function  $ f(R,T) $ have been proposed for cosmology \cite{36}. They are (a) $R + 2f(T) $ (b) $ f_{1}(R) + f_{2}(T)$ and (c) $ f_{1}(R) + f_{2}(R)f_{3}(T).$ The purpose of this work is to model a universe in f(R,T) gravity which meets observational constraints \cite{6} \& \cite{67}. For this, we consider the first simple alternative of  $f(R,T)$ as  $f(R,T) = R + 2 \lambda T$, where $\lambda$ is an arbitrary constant to comply our objective.\\

We solve  $ f(R,T) $ gravity field equations (\ref{4} ) for Bianchi-I anisotropic spacetime given by
\begin{equation}
	\label{6}
	ds^2 = dt^2-a^2dx^2-b^2dy^2-c^2dz^2,
\end{equation}
where a, b and c are scale factors along spatial directions and they depend on time only. We have taken $ L_m = - p $ and  $ f(R,T) = R + 2\lambda T $,  so that we get $T_{ij}$ as
\begin{equation}\label{7}
	T_{ij}= (\rho + p) u_i u_j - p g_{ij}.
\end{equation}

We get the following field equations:
\begin{equation}
	\label{8}
	\frac{\ddot{b}}{b}+\frac{\ddot{c}}{c}+\frac{\dot{b}\dot{c}}{bc}=-8 \pi ( p +  \mu  (3 p-\rho )),
\end{equation}
\begin{equation}\label{9}
	\frac{\ddot{a}}{a}+\frac{\ddot{c}}{c}+\frac{\dot{a}\dot{c}}{ac}= -8 \pi ( p +  \mu  (3 p-\rho )),
\end{equation}
\begin{equation}\label{10}
	\frac{\ddot{a}}{a}+\frac{\ddot{b}}{b}+\frac{\dot{a}\dot{b}}{ab}=-8 \pi ( p +  \mu  (3 p-\rho )),
\end{equation}
and
\begin{equation}\label{11}
	\frac{\dot{a}\dot{b}}{ab}+\frac{\dot{b}\dot{c}}{bc}+\frac{\dot{c}\dot{a}}{ac}=8 \pi  ( \rho +  \mu  (3 \rho -p)).
\end{equation}

From Eqs.(\ref{8})$-$(\ref{10}), we get
\begin{equation}\label{12}
	\frac{\ddot{b}}{b}+\frac{\ddot{c}}{c}+\frac{2\dot{b}\dot{c}}{bc} =2\frac{\ddot{a}}{a}+\left(\frac{\dot{b}}{b}+\frac{\dot{c}}{c}\right)\frac{\dot{a}}{a},
\end{equation}
which is re-written as:
\begin{equation}\label{13}
	\frac{d}{dt}\left(\frac{\dot{(bc)}}{bc}\right)+\left(\frac{\dot{(bc)}}{bc}\right)^{2}=2\frac{d}{dt}\left(\frac{\dot{a}}{a}\right)+2\frac{a^{2}_{4}}{a^{2}}+\frac{\dot{a}\dot{(bc)}}{abc}.
\end{equation}
This equation gives a particular solution
\begin{equation}\label{14}
	\frac{\dot{(bc)}}{bc}-\frac{2\dot{a}}{a}=0.
\end{equation}
 So we get a following relationship among the metric coefficients
\begin{equation}\label{15}
	a^2=bc.
\end{equation}
Therefore we may assume
\begin{equation}\label{16}
	b =ad,\; c = \frac{a}{d}\; d=d(t).
\end{equation}
With these choices of metric coefficients, the f(R,T) field equations take the following form:
\begin{equation}\label{17}
	2{\left(\frac{\ddot{a}}{a}\right)}+\left(\frac{\dot{a}}{a} \right)^2  =
 -8 \pi  ( p +  \mu  (3 p-\rho ))-{\left( \frac{\dot{d}}{d}\right)}^2 ,
\end{equation}
\begin{equation}\label{18}
	3\left(\frac{\dot{a}}{a} \right)^2 = 8 \pi  ( \rho +  \mu  (3 \rho -p))+ {\left( \frac{\dot{d}}{d}\right)}^2,
\end{equation}
\begin{equation}\label{19}
	\frac{\dot{d}}{d}= \frac{k}{a^3}.
\end{equation}
 Field equations  (\ref{17}) and (\ref{18}) are General relatively Bianchi-I field equations for $\mu =0$ which when solved, give decelerating universe.  $\mu$ terms are present due to f(R,T)= R + 8 $\pi \mu$ T. We  must expect acceleration ie $q\leq 0$ at present due to their presence . So what we do, we name  $p_\mu$, the  $\mu$ terms with p in Eq. (\ref{17}) and  $\rho_\mu$, the $\mu$ terms with $\rho$ in Eq. (\ref{18}). There is term  $\frac{\dot{d}}{d}^2$ in both these equations , which appeared due to anisotropy, so we give them name $p_{\sigma}$ and $\rho_{\sigma}$. Accordingly we define:
\begin{equation}\label{20}
	p_{\mu} =\mu  (3 p-\rho ), ~ \rho_{\mu} =  \mu  (3 \rho -p),~ p_{\sigma}=\rho_{\sigma}~=  ~\frac{\dot{d}}{d}^2=\frac{k^2}{a^6}.
\end{equation}

	If we look at field equations (\ref{17}) and (\ref{18}), we find that due to presence of terms $\mu  (3 \rho -p)$ in these equations, the conservation of energy momentum tensor $ T^{ij}_{;j}=0$ which is calculated as
	$ \dot{\rho} + 3\rho(p+\rho)=0$ does not satisfy in our model.  This can also be seen from field equations (\ref{4}) of f(R,T) gravity, in which the whole right hand side will be conserved which contain so many other terms apart from $T^{ij}$. However, as an alternative to the energy conservation equation, we obtain the following equation in our model: $(1+3\mu) \dot{\rho} - \mu \dot{p} = - 3 (1+2\mu) H (p +\rho).$
	It is clear from this that in absence of $\mu$ term, we get energy conservation law.\\

We note that $\rho_\mu$ and $p_\mu$ are contributions of  f(R,T) in pressure and density and others are contribution og anisotropy in the universe. So we may define energy parameters  $ \Omega _m$, $\Omega _{\mu}$, $\Omega _{\sigma}$ and equations of state parameters $\omega$ and as $\omega_\mu$ as follows;
\begin{equation}\label{21}
	\Omega _m=\frac{8\pi\rho}{3H^2}, ~  \Omega _{\mu}=\frac{8\pi \rho_\mu}{3H^2},~ \Omega _{\sigma}=\frac{8\pi \rho_\sigma}{3H^2}, ~ \omega_m =\frac{p}{\rho},~ \omega_\mu =\frac{p_\mu}{\rho_\mu} .
\end{equation}
Here suffixes m and $\mu$ stands for parameters of baryon matter and f(R,T) effect. We may also interpreted parameters with $\mu$  suffix as turns arriving due to curvature dominance of $f(R,T)$ gravity. We note that $\omega_{\sigma} =\frac{p_{\sigma}}{\rho_{\sigma}}=1$ \\

	Field   Eqs. (\ref{17}) and (\ref{18}) are described as:
\begin{equation}\label{22}
	H^2 (1-2 q)=-8 \pi  ( p +  p_{\mu} + p_{\sigma}  )  , ~~ 	3 H^2=8 \pi  ( \rho + \rho_\mu +  \rho_{\sigma} ),~~ \Omega _m +	 \Omega _{\mu} +\Omega_{\sigma}=1,
\end{equation}
and
\begin{equation}\label{23}
	\omega _{\mu }=\frac{3 \omega _m-1}{3-\omega _m} ~~~~ \omega _m = \frac{3 \omega _{\mu }+1}{\omega _{\mu }+3}.	
\end{equation}
From Eqs.  (\ref{20}) and (\ref{22}), we derive expressions for 
 p,   $p_{\mu},\rho~ \text{and}~ \rho_\mu$ as follows:

  \begin{multline}
  	     \\
         8 \pi p= \frac{H^2 (2 (3 \mu +1) q-1)}{(4 \mu +1) (2 \mu +1)} -  \frac{k^2}{(1+2\mu) a^6} ,\\
         8\pi\rho =\frac{H^2 (2 \mu  (q+4)+3)}{(4 \mu +1) (2 \mu +1)} - \frac{k^2}{(1+2\mu) a^6},\\
         8 \pi\rho_{\mu} = \frac{2 H^{2} \mu(12 \mu -q+5)}{(4 \mu +1) (2 \mu +1)} - \frac{2 k^2 \mu}{(1+2\mu) a^6},\\ 
         \label{24}
         8\pi p_{\mu} = \frac{2 H^{2} \mu(- 4 \mu +(8 \mu +3) q-3)}{(4 \mu +1) (2 \mu +1)} - \frac{2 k^2 \mu}{(1+2\mu)a^6}.\\  
     \end{multline}	
 
 
\section{Hubble, Deceleration and Jerk parameters:}
The jerk and snap are related with third and forth order of derivatives of scale factor. They are used in examining instability of a cosmological model. jerk is also used in statefinder diagnostic. their definition is as follows: jerk $j=\frac{\dddot{a}}{a H^3}$ and snap $s= -\frac{\ddddot{a}}{a H^4}. $
The jerk, snap, deceleration and Hubble parameters are associated to each other through the followings Eqs.:
\begin{equation}\label{25}
		H_z(1+z)=(q+1)H	,
\end{equation}
\begin{equation}{\label{26}}
	j(z) = q(z) +2 {q(z)}^2 + (1+z) \frac{d q(z)}{d z},
\end{equation}
\begin{equation}{\label{27}}
	s(z) = ( 3 q(z) +2 ) j(z) + \frac{d j(z)}{d z} (1+z).
\end{equation} 
In $\Lambda$ CDM model $j=1$ and other models where jerk is variables, it varies very slowly. For instance, in \cite{66}, jerk vary from $0.99 -1.04$, so we model an Universe with constant jerk parameter in $f(R,T)$ gravity.
 so we solve Eq. (\ref{26}) for constant jerk i.e. we take $j$ = constant.
 We obtain following expression for deceleration parameter.
 \begin{equation}\label{28}
 	q =	\frac{1}{4}\left(-1+\sqrt{-8 j-1}\text{Tan}\left[\frac{1}{2}\left(-2 \tan ^{-1}\left(\frac{4 \sqrt{-8 j-1} q_0+\sqrt{-8 j-1}}{8 j+1}\right)-\sqrt{-8 j-1} \log (z+1)\right).\right]\right),
 \end{equation} 
 where we have used $q = q_0~ \text{at~ present}~ i.e.~ z=0.$
 The deceleration parameter $q$ is related to Hubble parameter through the following differential equation.
 \begin{equation}\label{29}
 	H_z(1+z)=(q+1)H.
 \end{equation}
 So, using  Eq. (\ref{28}) and integrating Eq. (\ref{29}), we may get the  expression for Hubble parameter, 
 \begin{equation}\label{30}
 	H = H_0 e^{\int_{0}^{z}\frac{(q+1)dz}{(1+z)}}.
 \end{equation}
 We note that the Eq. (\ref{30}) has three unknown parameters $H_0,~ j~ \text{and}~ q_0.$
 
\section{Fitting  46 OHD in the Model to Evaluate Hubble, Decelerating and Jerk Parameters:}
We use a set of 46 Hubble's observed data commonly known as OHD (Observed Hubble data set) which consist of empirical values of Hubble constant at different red shift in the range $0\leq z\leq 2.36$ along with errors in each value in form of standard deviation (\cite{68}) $-$ (\cite{82}). From these and Eq. \ref{30} we  estimate statistically  $H_0,~ j~ \text{and}~ q_0$ on the basis of minimum chi squire given by 
\begin{equation}\label{31}
	\chi^{2}( H_0,j,q_0) =\frac{1}{46} \sum\limits_{i=1}^{46}\frac{[Hth(z_{i},H_0,j,q_0) - H_{ob}(z_{i})]^{2}}{\sigma {(z_{i})}^{2}}.
\end{equation}
We find that $ H_0 = 68.95,~j=0.95~ \text{and}~ q_0 = -0.57$  at minimum  $\chi2$ =0.5378. It is a good fit.  we rewrite Eq. (\ref{28}) and integrate Eq. (\ref{30}) by taking $q_0=-0.57$ and $j=0.95$ which are estimated values as per OHD data set. We get the following expressions for deceleration and Hubble parameters as:
\begin{equation}\label{31a}
	q = \frac{0.483144 \left( (z+1)^{2.93258}-5.18713\right)}{(z+1)^{2.93258}+2.5491},	
\end{equation}
and
\begin{equation}\label{31b}
	H = 0.54 H_0 (1+z)^{0.017}(3.55 + z(3+z(3+z)))^{0.489}.
\end{equation}

We present  the following figure to show the nature of parameters $H$ and $q$ graphically and how they respond to the observational data set.
 Figure $1(a)$ shows the growth of deceleration parameter $`q'$  over red shift $ `z'$. It describes that in the past the universe was decelerating. At transition redshift $z_t$=0.7534, where $q\sim0$, it changed its behavior and started accelerating.
Figures $1(b)$ and $1(c)$ describe the growth of Hubble parameter $H$ and  expansion rate $H/(1+z) = \dot{a}/a_0$ over red shift $ `z'$ respectively. Hubble parameter is increasing function of red shift which means that in the past Hubble constant was more. It is gradually decreasing over time. Expansion is high at present which indicates that universe is accelerating. These figures also show that theoretical graph passes near by through the dots which are observed values at different red shifts. Vertical lines are error bars. Figure $1(d)$ is likelihood probability curve for Hubble parameter. Estimated value $H_0=68.95$ is at the peak. Figures $1(e)$, $1(f)$, and $1(g)$ are contour 1$\sigma$, 2$\sigma$ and 3$\sigma$ region plots for the estimated values and $\chi^2$.

\begin{figure}[H]
	(a)	\includegraphics[width=9cm,height=8cm,angle=0]{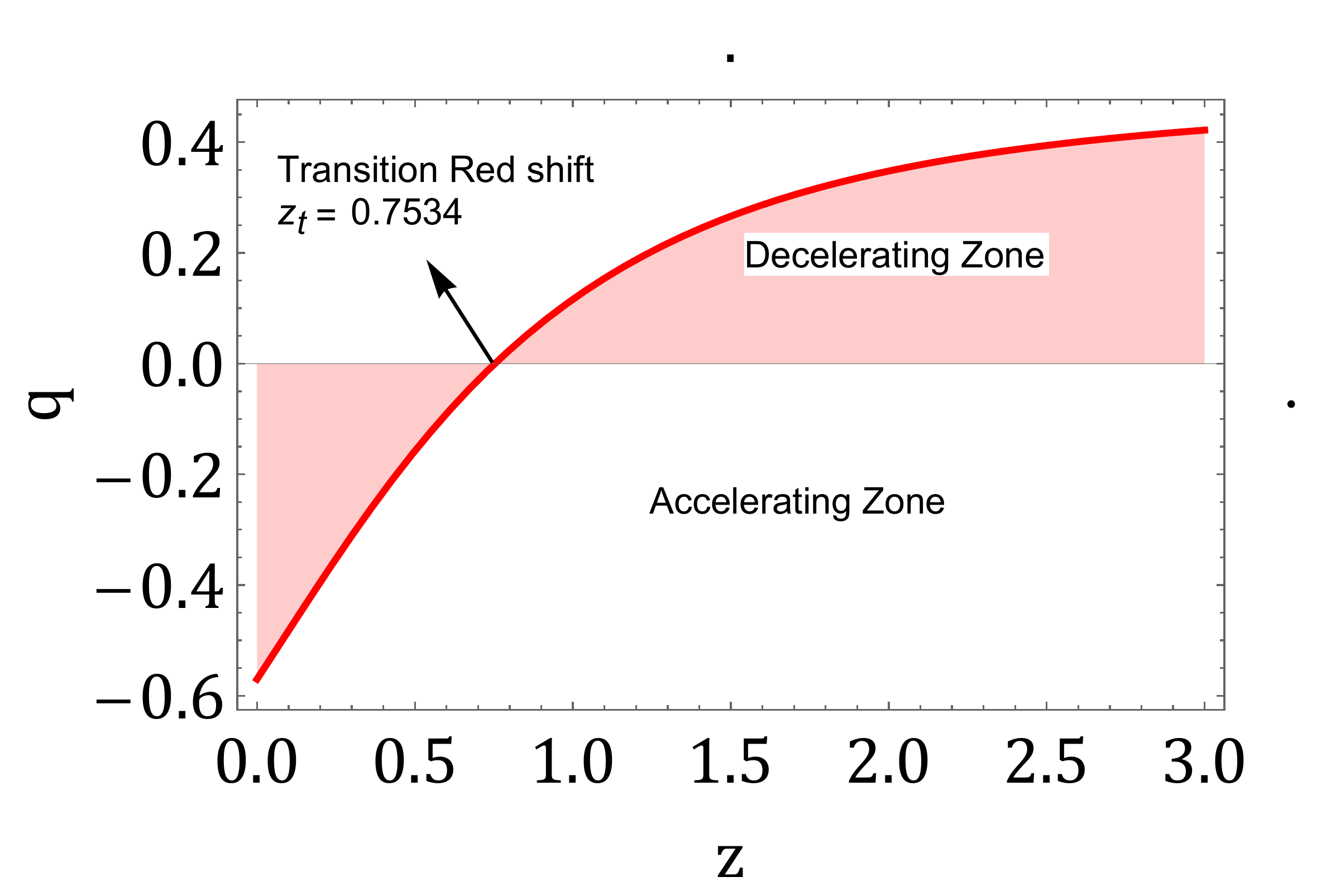}
	(b)	\includegraphics[width=9cm,height=8cm,angle=0]{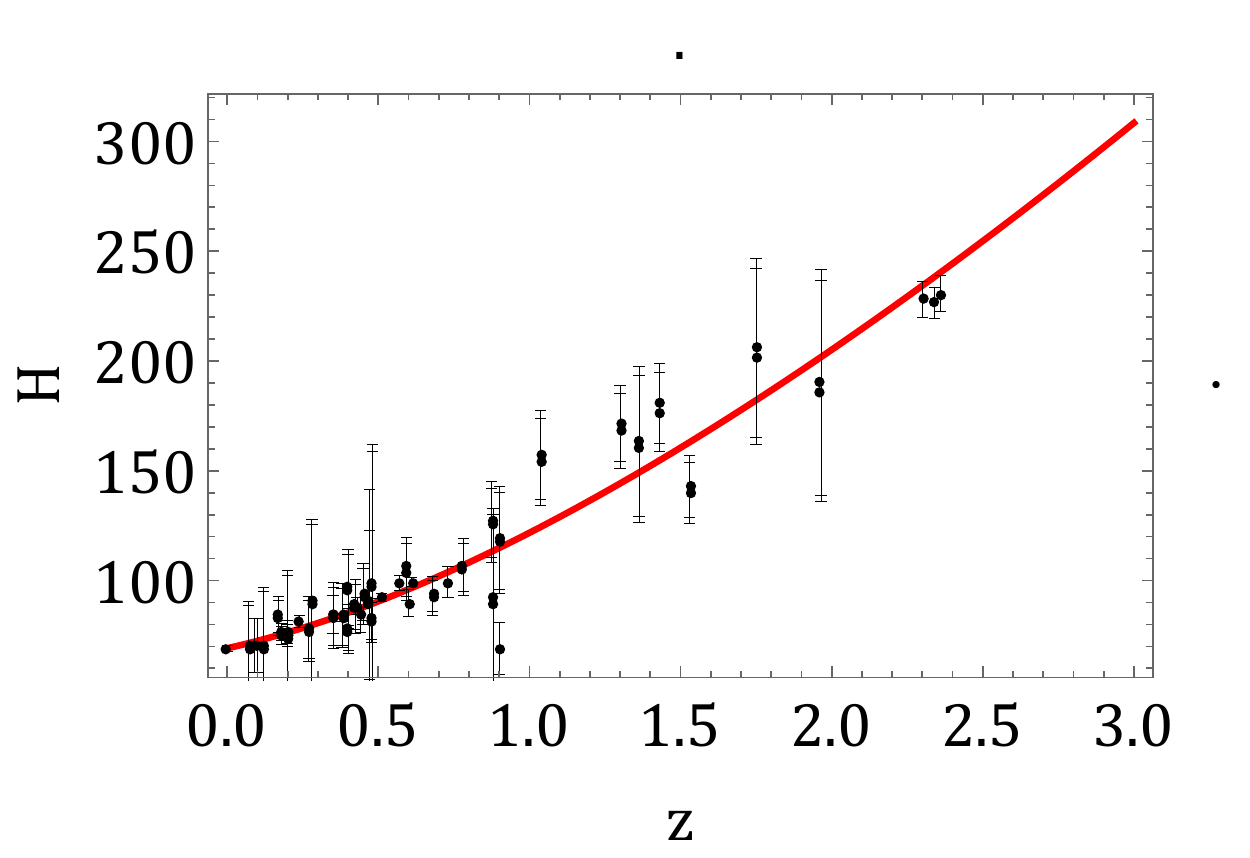}
    (c) \includegraphics[width=9cm,height=8cm,angle=0]{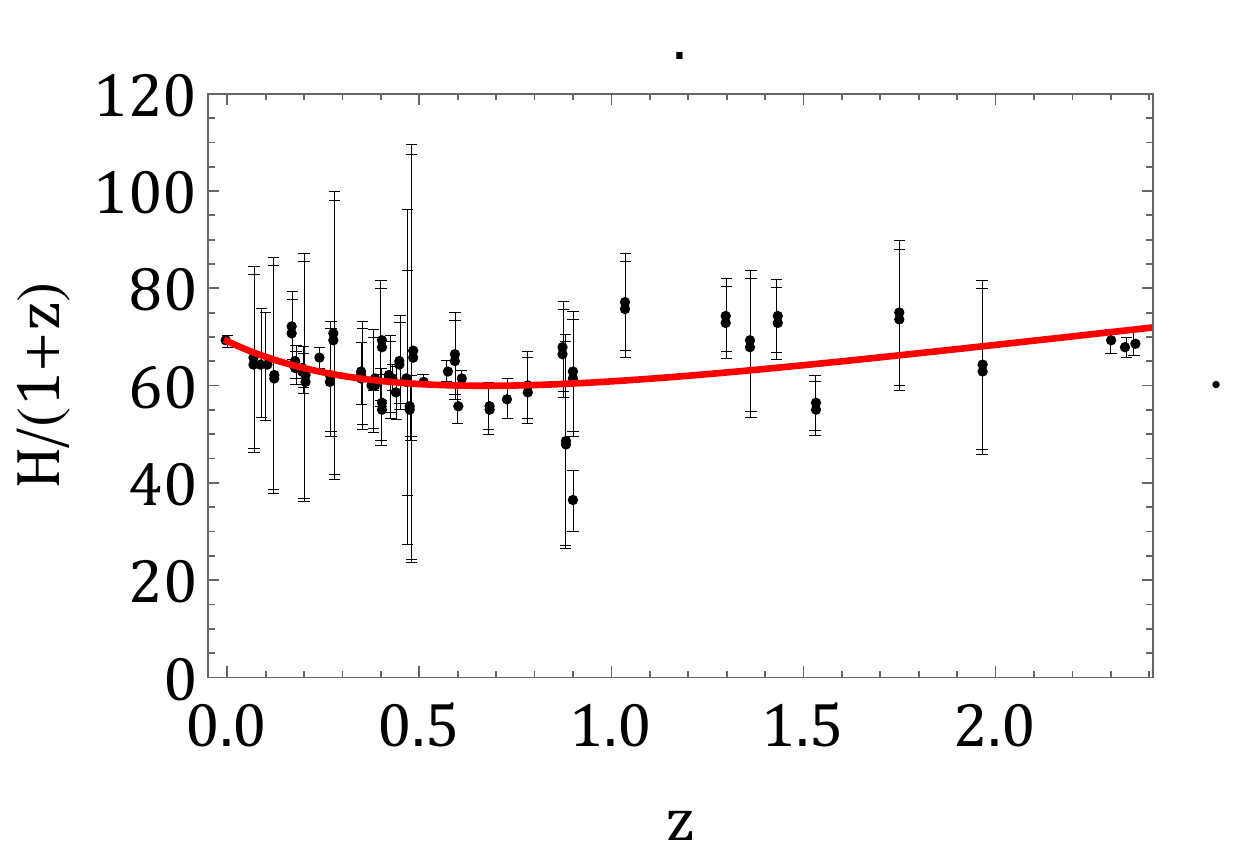}
	(d) \includegraphics[width=9cm,height=8cm,angle=0]{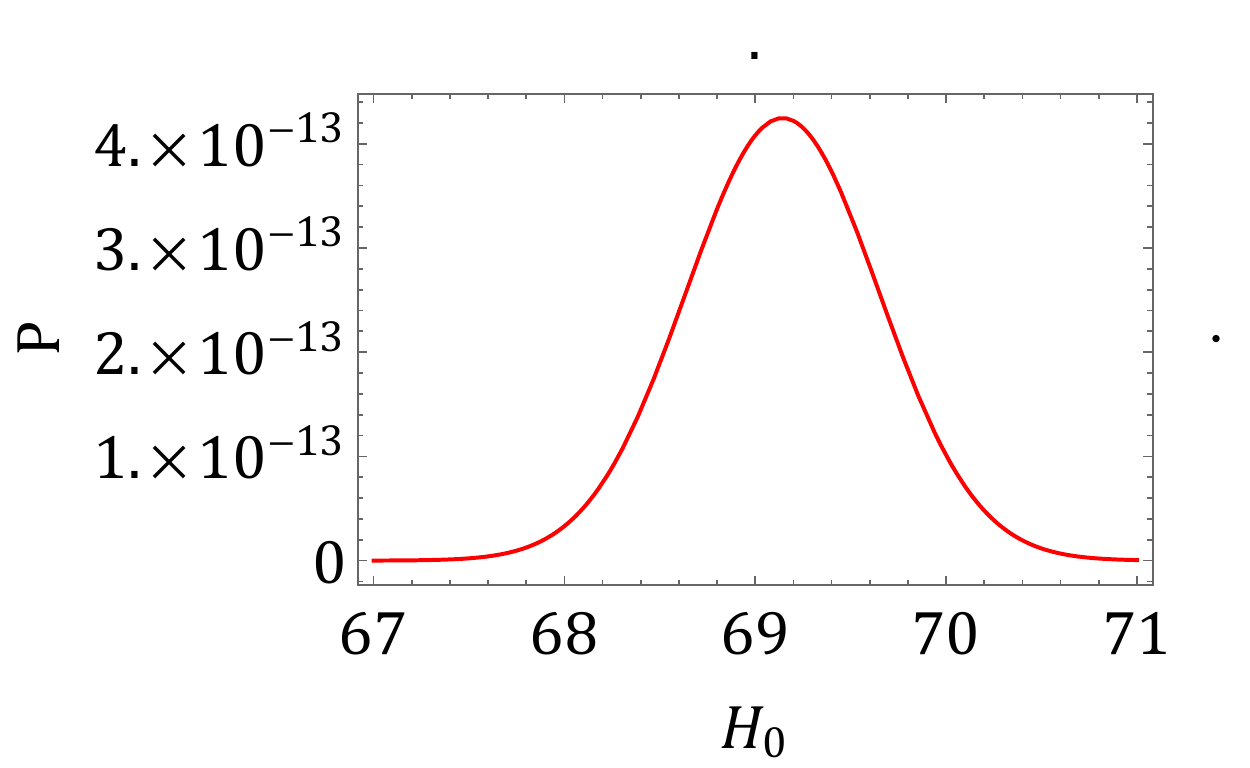}
	(e)	\includegraphics[width=9cm,height=8cm,angle=0]{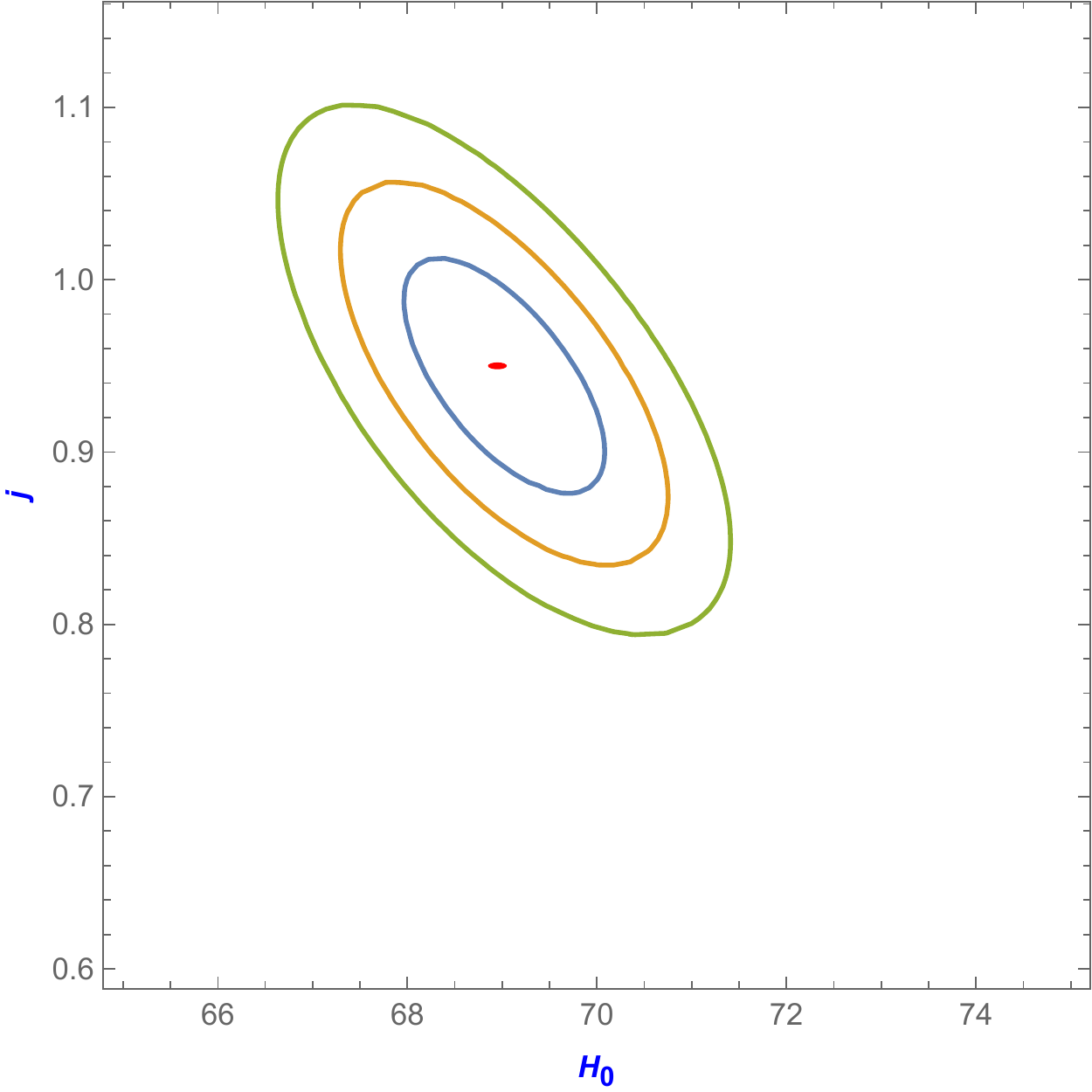}
	(f) \includegraphics[width=9cm,height=8cm,angle=0]{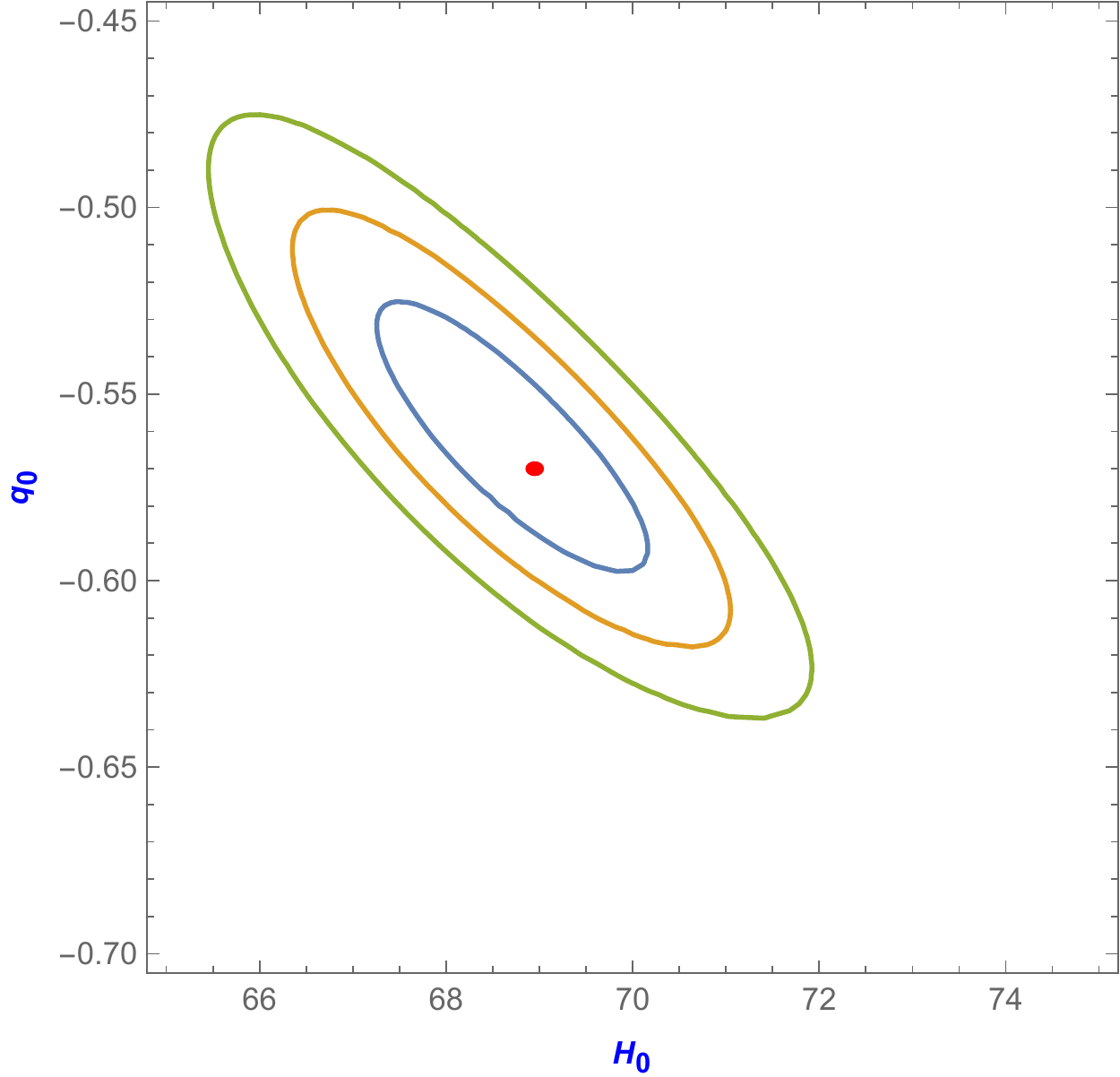}
\end{figure}

\begin{figure}
    (g) \includegraphics[width=9cm,height=8cm,angle=0]{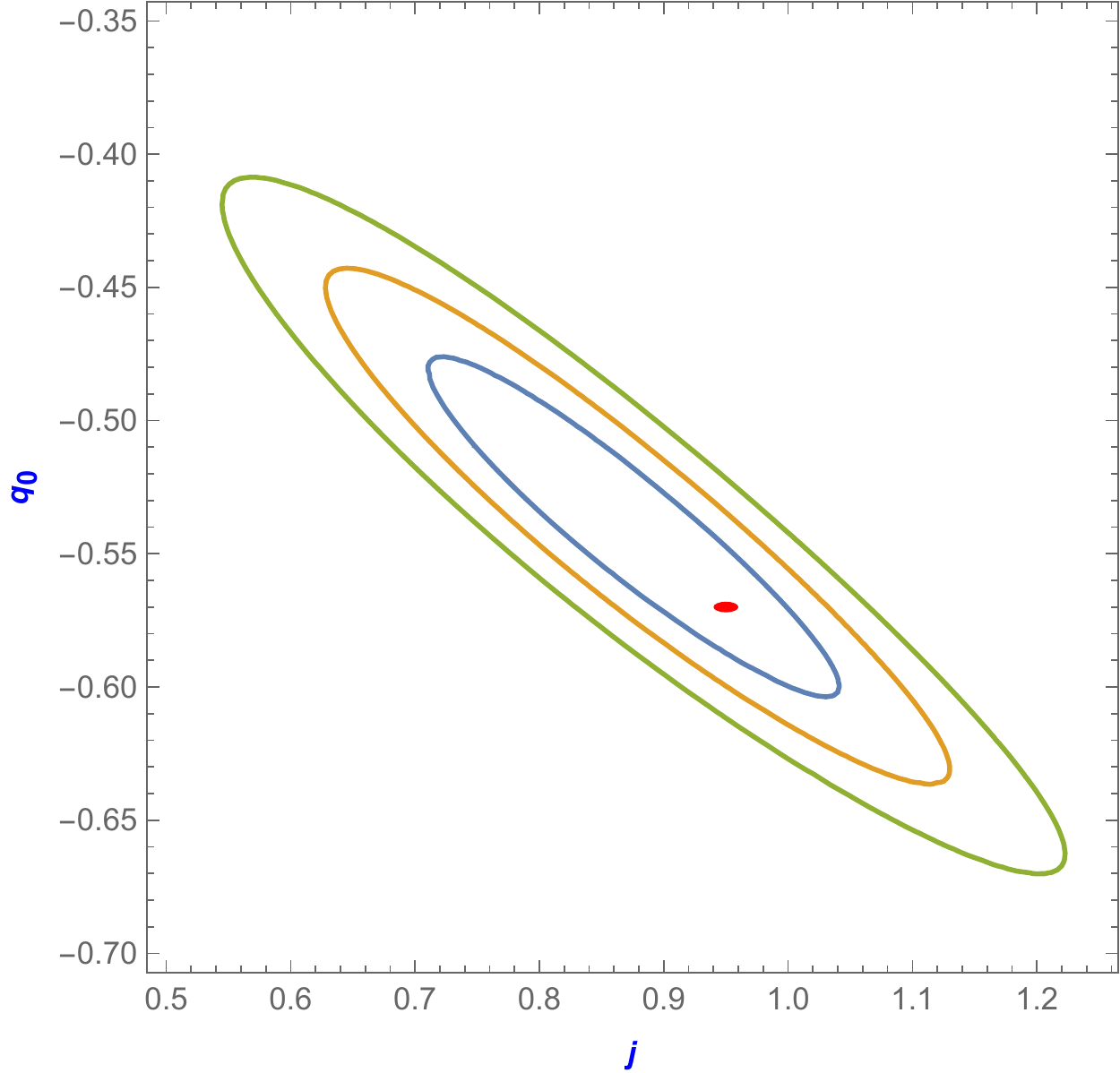}
	\caption{ Figure $(a)$ shows the growth of deceleration parameter $`q'$  over red shift $ `z'$. It describes that in the past the universe was decelerating. At transition redshift $z_t$=0.7534, where $q\sim0$, it changed its behavior and started accelerating.  Here $q_0 =-0.57~\text{and}~ j = 0.95.$  Figures $(b)$ and $(c)$ are the error bar plots  for  Hubble parameter $H$ and  expansion rate $H/(1+z) = \dot{a}/a_0$ over red shift $ `z'$ respectively. Figure $(d)$ is likelihood probability curve for Hubble parameter. Estimated value $H_0=68.95$ is at the peak. Figures $(e)$, $(f)$, and $(g)$ are 1$\sigma$, 2$\sigma$ and 3$\sigma$  contour  region plots for the estimated values and $\chi^2$.}
   \end{figure}

\section{$H_0$ from Union 2.1 compilation and Pantheon Pan-STARRS1 Data Set:}
In this section we use SNIa 580 data sets of distance modulus $ \mu(z) $ from
 Union 2.1 compilation \cite{83} and $66$ Pantheon apparent magnitude $m_b$ data's consisting of the latest compilation of SN Ia $40$ bined plus 26 high redshift  data's in the  range $0.014 \leq z \leq 2.26. $ \cite{84,85,86}.  \\
 
The distance modulus (D.M.)$ \mu(z) $ is defined as
\begin{equation}\label{32}
	\mu(z)= m_b - M = 5 Log d_l(z)+\mu_{0},
\end{equation}
where $ m_b $ and $ M $ are the apparent and absolute magnitude of the standard candle respectively. The luminosity distance $ D_l(z) $  and nuisance parameter
$ \mu_0 $ are defined by 
\begin{equation}\label{33}
	\mu_0= 25+5 Log \big(\frac{c}{H_0} \big),   
\end{equation}
and
\begin{equation}\label{34}
	d_l(z)=(1+z)H_0 \int_0^z{\frac{1}{H(z*)} dz*},
\end{equation}
respectively.\\

As the the absolute magnitudes of a standard candles are more or less same.  Its value is obtained as $ M = -19.09$\cite{13}. So, we can obtain apparent magnitude (M.B.) $m_b$ as follows:
\begin{equation}\label{35}
	m_b =  5 Log[ (1+z) c \int_0^z{\frac{1}{H(z*)} dz*}]+ 5.91.
\end{equation}
For our model, Eqs.  (\ref{32}) and (\ref{35}) will be read as
  \begin{equation}\label{36}
  	\mu(z) = 5 Log (1+z)H_0 \int_0^z{\frac{1}{ H_0\left(0.3 (z*+1)^3+0.7\right)^{0.5}} dz*}+ 25+5 Log \big(\frac{c}{H_0} \big),
  \end{equation}
and 
\begin{equation}\label{37}
	m_b =  5 Log[ (1+z) c \int_0^z{\frac{1}{ H_0\left(0.3 (z*+1)^3+0.7\right)^{0.5}} dz*}]+ 5.91	.
\end{equation}
As these equations contain $H_0$, so we can estimate this in the same manner as in section $4$, by using following expressions for $\chi$ squires.
\begin{equation}\label{38}
	\chi^{2}(H_{0}) = \frac{1}{579}  \sum\limits_{i=1}^{580}\frac{[\mu th (z_{i},H_{0}) - \mu ob(z_{i})]^{2}}{\sigma {(z_{i})}^{2}},
\end{equation}
and 
\begin{equation}\label{39}
	\chi^{2}(H_{0}) =\frac{1}{65}  \sum\limits_{i=1}^{66}\frac{[m_{b} th (z_{i},H_{0}) - m_{b} ob(z_{i})]^{2}}{\sigma {(z_{i})}^{2}}.
\end{equation}
It is found that  $\{H_0\to 69.7171\}$ at $\chi^2= \{0.97323\}$ for SNIa 580 distance modulus $\mu$ data's and $\{H_0\to 78.8906\}$ at $\chi^2= \{1.1144\}$ for 66 apparent magnitude Pantheon data's. We construct one more data set  of $657$ data's  by combining Hubble $77$ data set with SNIa 580 distance modulus $\mu$ data's and use following $\chi$ squires
\begin{equation}\label{40}
	\chi^{2}(H_{0}) = \frac{1}{579}  \sum\limits_{i=1}^{580}\frac{[\mu th (z_{i},H_{0}) - \mu ob(z_{i})]^{2}}{\sigma {(z_{i})}^{2}} +\frac{1}{76} \sum\limits_{i=1}^{77}\frac{[Hth(z_{i},H_0) - H_{ob}(z_{i})]^{2}}{\sigma {(z_{i})}^{2}}.
\end{equation}
  This gives the estimated value of $H_0$ as $H_0 = 69.367$ at minimum $\chi2 = 0.864739$. Same way we construct two more by joining ~D.M. $\mu$ and  A.M. $m_b$ and by joining all the three $OHD$, D.M. $\mu$ and A.M $m_b$.\\
  We present all our statistically evaluated findings in the following Table 1:
  
  \begin{table}[H]
  	\caption{ Statistical estimations for Hubble parameter $H_0$}
           \begin{tabular}{|c|c|c|c|c|c|c|c|}
  			\hline
  			\\
  			Datasets &~	 $H_{0}$~ &~   $\chi^2$  ~&	Datasets &~	 $H_{0}$~ &~   $\chi^2$  ~ \\
  			\\	
  			\hline
  			\\
  			$OHD$  &~  $ 68.95 $ ~ &~ $0.5378~ $  & ~	D.M	$\mu$ &~  $69.7171$	 ~&~ $0.97323~ $\\
  			\\	
  			\hline
  			\\
  			A.M	$m_b$ &~  $78.8906$	 ~ &~ $1.1144~ $&~	$OHD$ +	D.M	$\mu$ &~  $69.367$	 ~ &~  $ 0.864739~$\\
  			\\	
  			\hline
  			\\
  		    ~D.M	$\mu$ + A.M	$m_b$ &~  $78.5398$	  ~&~ $ 1.10473~$&~	$OHD$ + D.M	$\mu$ + A.M	$m_b$	 &~  $77.914$	 ~&~  $ 1.04471~$\\
  		\hline			
  		\end{tabular}
        \end{table}
   We present Figure $2$ to display our findings graphically. Figs. $2 (a)$ and $2(b)$ are the error bar plot and likelihood probability curve for  distant modulus $\mu$ and Hubble parameter respectively over red shift $ `z'$. Figs. $2(c)$ and  $2(d)$ are the error bar plot and likelihood probability curve for apparent magnitude and Hubble parameter over red shift $ `z'$. Estimated values $H_0=69.7171~ \text{and}~  78.8906$ are at the peaks in the two likelihood curves.
   
    \begin{figure}[H]
    	(a)	\includegraphics[width=9cm,height=8cm,angle=0]{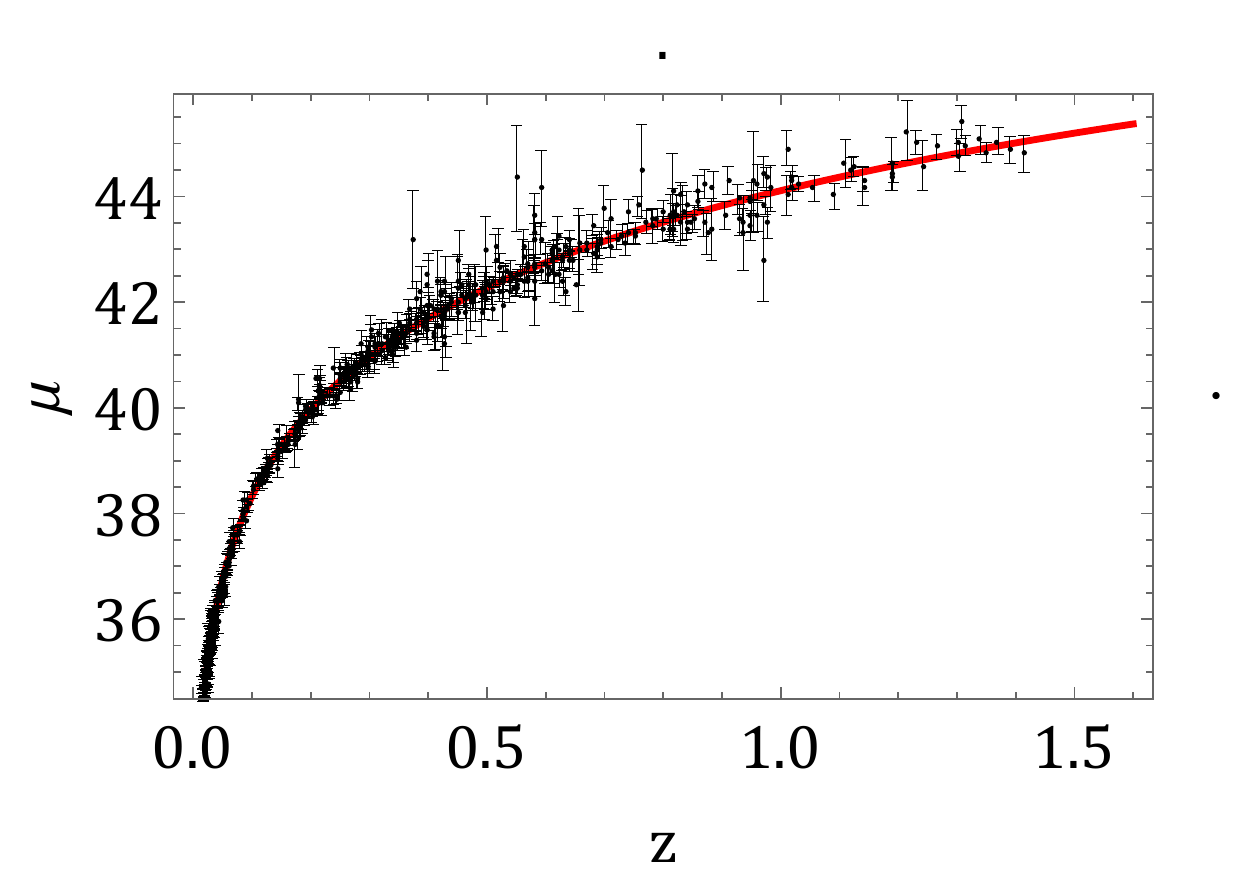}
    	(b) \includegraphics[width=9cm,height=8cm,angle=0]{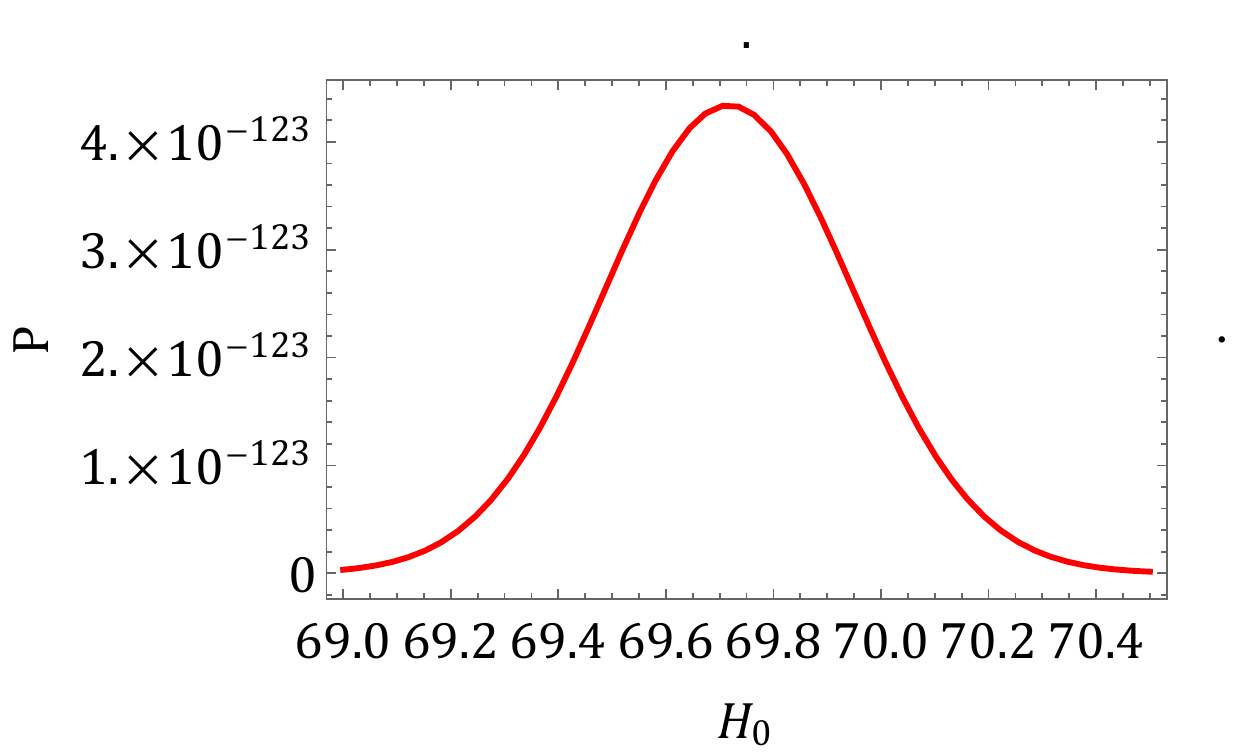}
    	(c) \includegraphics[width=9cm,height=8cm,angle=0]{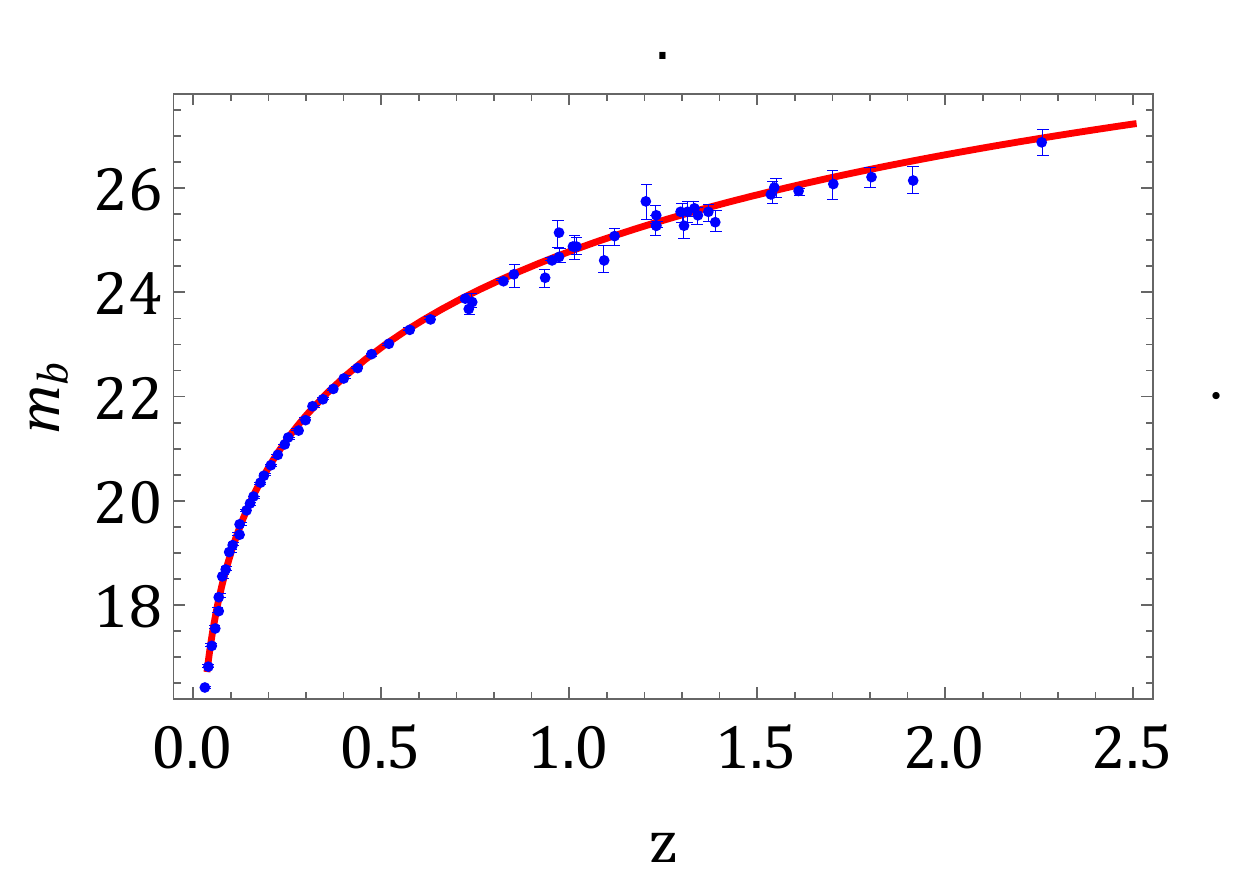}
    	(d)	\includegraphics[width=9cm,height=8cm,angle=0]{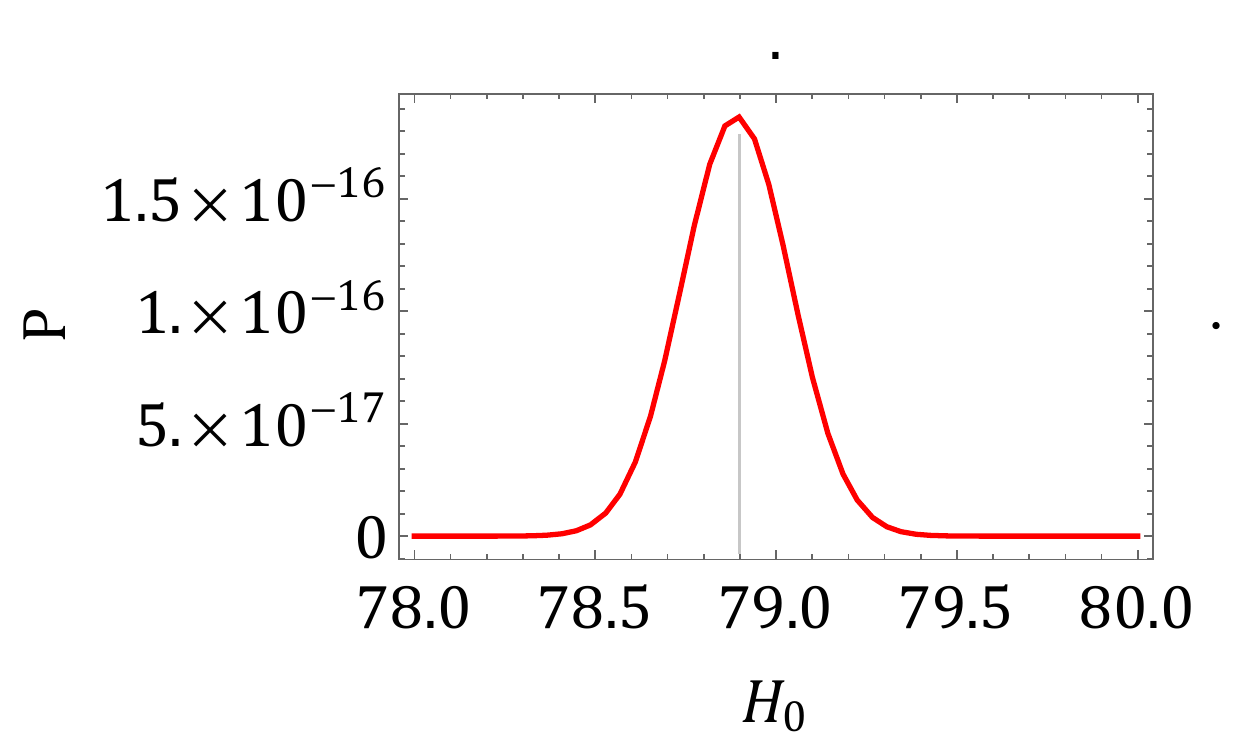}
    \caption{ Figs. $2(a)$ and $2(b)$ are the error bar plots and likelihood probability curve for  distant modulus $\mu$ and Hubble parameter respectively over red shift $ `z'$. Figures $2(c)$ and $2(d)$ are the error bar plot and likelihood probability curve for apparent magnitude and Hubble parameter over red shift $ `z'$. Estimated values $H_0=69.7171~ \text{and}~  78.8906$ are at the peaks in the two likelihood curves.}
    \end{figure}

\section{Estimation of $\mu$ on the basis of Energy Parameters:}
From Eqs. (\ref{21}) and (\ref{24}), we obtain the following:
\begin{equation}\label{41}
\Omega_{m0}	+\frac{\Omega_{\sigma 0}}{1+2\mu}=\frac{2\mu (q_0 +4)+3}{3(2\mu +1)(4\mu)+1},
\end{equation} 
and
\begin{equation}\label{42}
\Omega_{\mu 0}	+\frac{2\mu\Omega_{\sigma 0}}{1+2\mu}=\frac{2\mu (12\mu -q_0+5 )}{3(2\mu +1)(4\mu)+1}.
\end{equation}
From these, we may confirm that $\Omega_{m 0}+\Omega_{\mu 0}+\Omega_{\sigma 0}=1$. Out of the three energy parameters, $\Omega_{\mu 0}$ is dominant, because we are getting acceleration in the universe due to this only. $\Omega_{\sigma 0}$ is contribution of anisotropy. We see that WMAP indicates very small temperature fluctuations in the universe, so its contribution is very minor. As is foreknown, $\Lambda$CDM model fits best on observation ground. It quotes the value  $\Omega_{m 0}= 0.296$. We consider same value for our model too. We take  $\Omega_{\sigma 0}=0.0005$,  so the value of  $\Omega_{\mu 0}$ comes to $0.6935$. From these values and using Eq. (\ref{41}), we get $\mu=  0.663073.$ 
\section{Density, Pressure and Equation of State parameter of the Universe:}
Having obtain the numerical values of the various model parameter, we can determine, baryon and $\mu$ densities, baryon and $\mu$ pressures and equation of state parameters $\Omega_m$ and $\Omega_\mu$. We rewrite Eq. (\ref{24}) as follows:
\begin{equation}\label{43}
	\rho =\frac{\rho _{\text{c0}} \Omega _{\text{m0}} \left((\frac{H}{H_0})^2 (2 \mu  (q+4)+3)-3 (4 \mu +1) (z+1)^6 \Omega _{\text{$\sigma $0}}\right)}{\left(2 \mu  \left(q_0+4\right)+3\right)-3 (4 \mu +1) \Omega _{\text{$\sigma $0}}},
\end{equation} 
\begin{equation}\label{44}
	\rho _{\mu }=\frac{  \rho _{\text{c0}} \Omega _{\text{$\mu $0}} \left((\frac{H}{H_0})^2 (12 \mu -q+5)-3 (4 \mu +1) (z+1)^6 \Omega _{\text{$\sigma $0}}\right)}{\left(12 \mu -q_0+5\right)-3 (4 \mu +1) \Omega _{\text{$\sigma $0}}},
\end{equation} 
\begin{equation}\label{45}
	p=\frac{\rho _{\text{c0}} \omega _{\text{m0}} \Omega _{\text{m0}} \left((\frac{H}{H_0})^2 (2 (3 \mu +1) q-1)-3 (4 \mu +1) (z+1)^6 \Omega _{\text{$\sigma $0}}\right)}{\left(2 (3 \mu +1) q_0-1\right)-3 (4 \mu +1) \Omega _{\text{$\sigma $0}}},
\end{equation}
\begin{equation}\label{46}
	p_{\mu }=\omega _{\text{$\mu $0}}\Omega _{\text{$\mu $0}}\rho _{\text{c0}}\left(\frac{(\frac{H}{H_0})^2 (-4 \mu +(8 \mu +3) q(z)-3)-3 (4 \mu +1) (z+1)^6 \Omega _{\text{$\sigma $0}}}{\left(-4 \mu +(8 \mu +3) q_0-3\right)-3 (4 \mu +1) \Omega _{\text{$\sigma $0}}}\right).
\end{equation}
  Eqs. (\ref{45}) and (\ref{46}) have unknown parameters $\omega _{\text{m0}} ~\text{and}~\omega _{\text{$\mu $0}}$. As at present baryon pressure is very much low, we take  $\omega _{\text{m0}}=0.0005 $ and acceleration in the universe requires negative $\mu$ pressure, so we take $\omega _{\text{$\mu $0}}=-1$. With these values, we can plot plots for densities and pressures which are given in the following figures:
   \begin{figure}[H]
  	(a)	\includegraphics[width=9cm,height=8cm,angle=0]{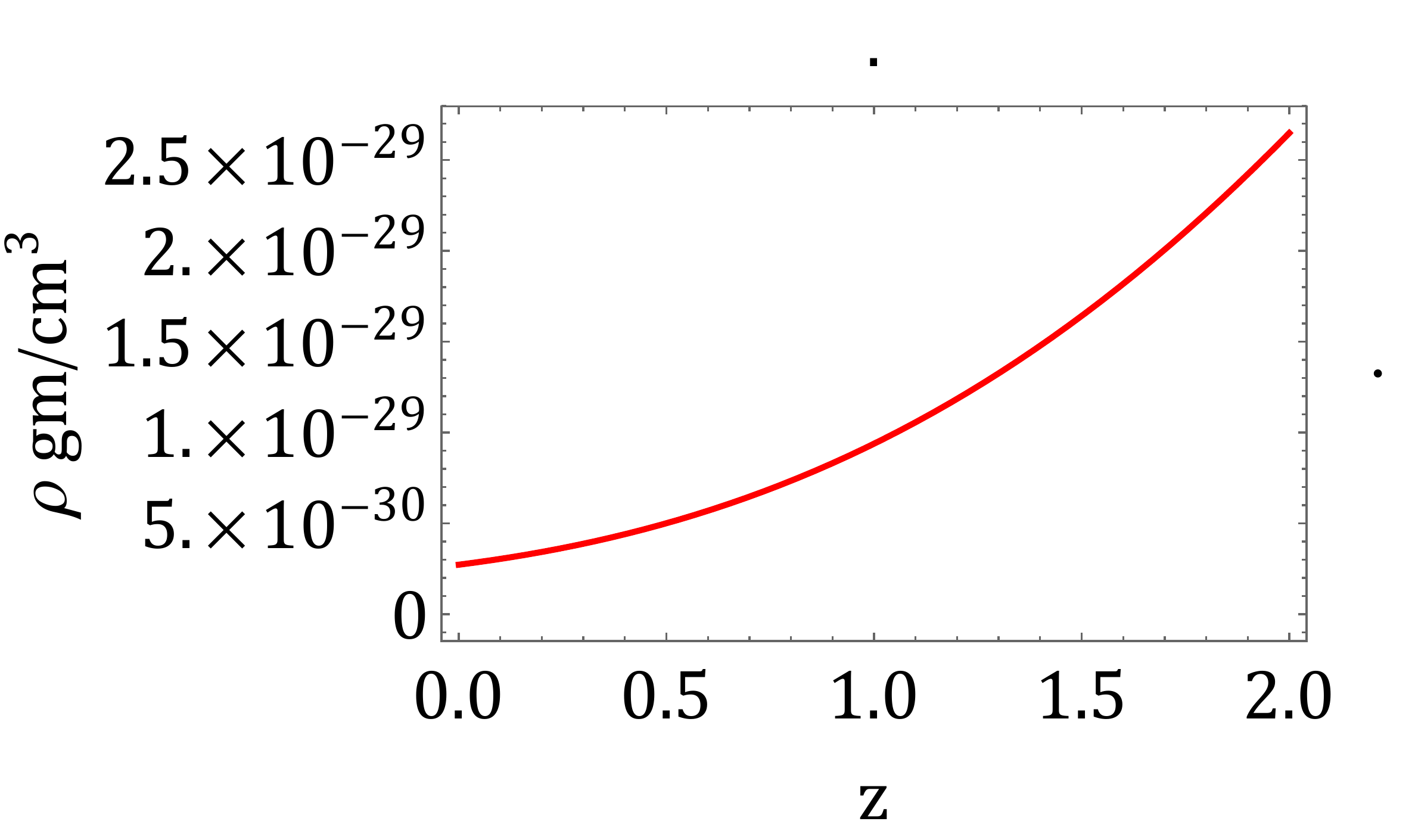}
  	(b) \includegraphics[width=9cm,height=8cm,angle=0]{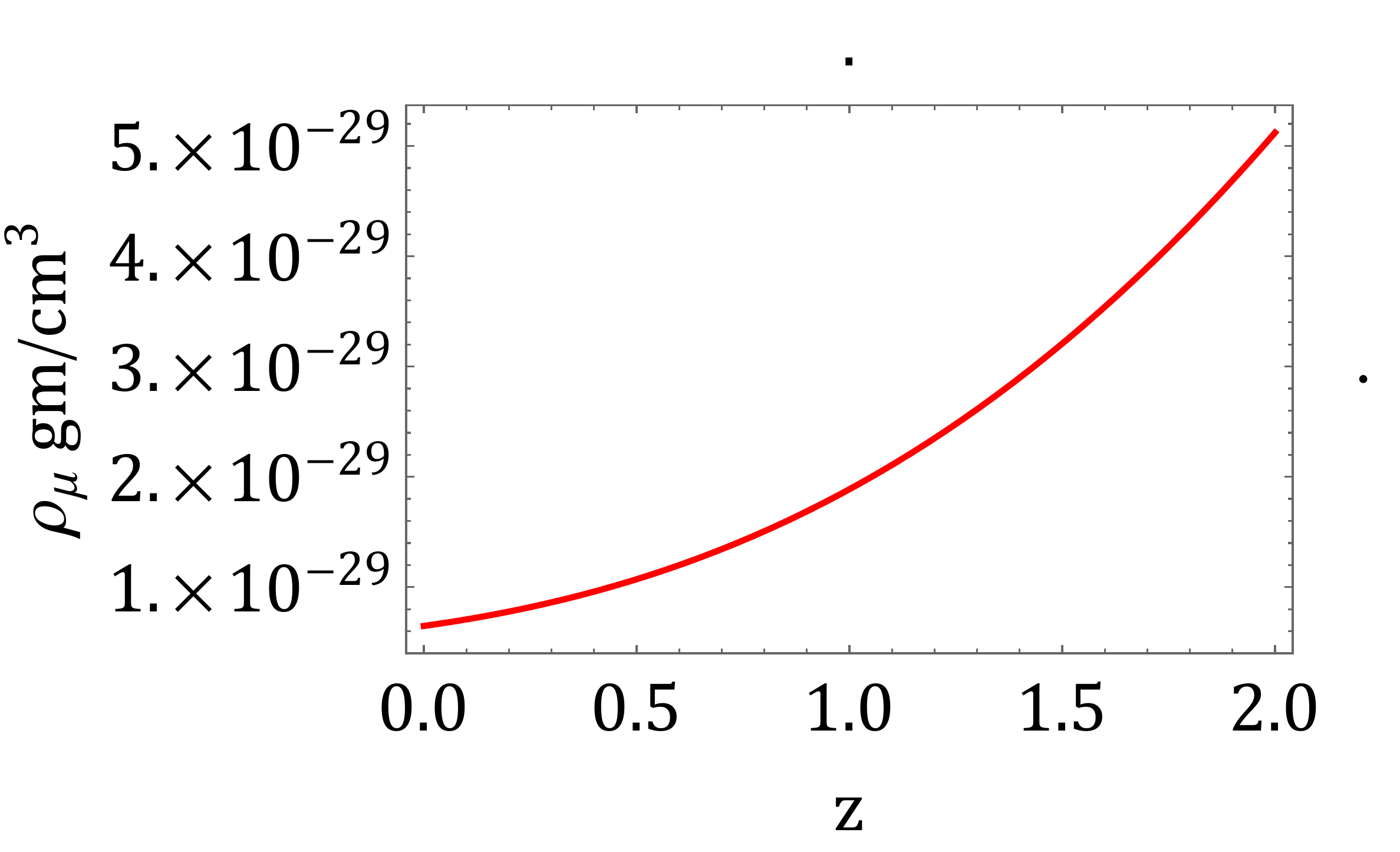}
  	(c) \includegraphics[width=9cm,height=8cm,angle=0]{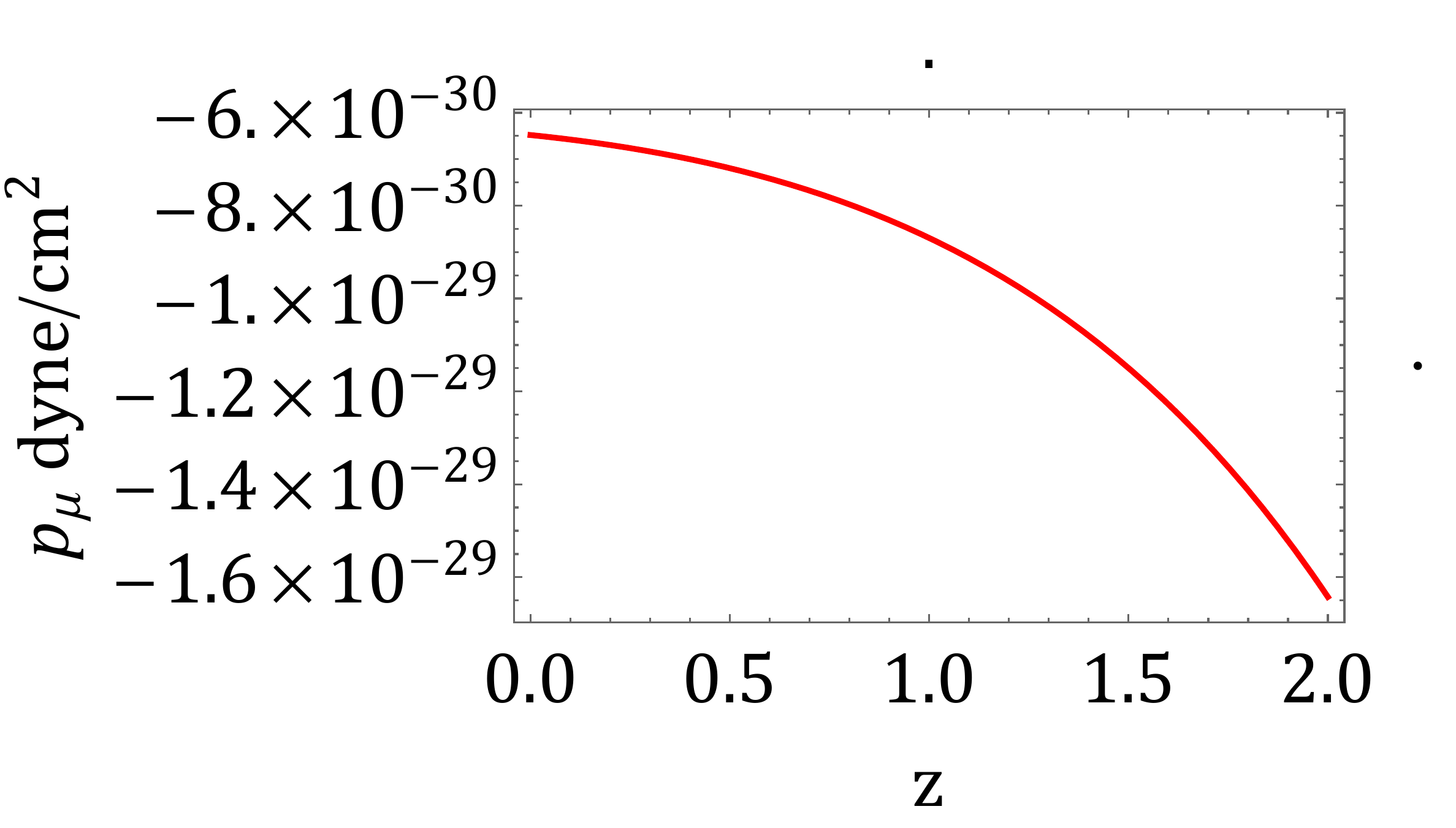}
  	(d)	\includegraphics[width=9cm,height=8cm,angle=0]{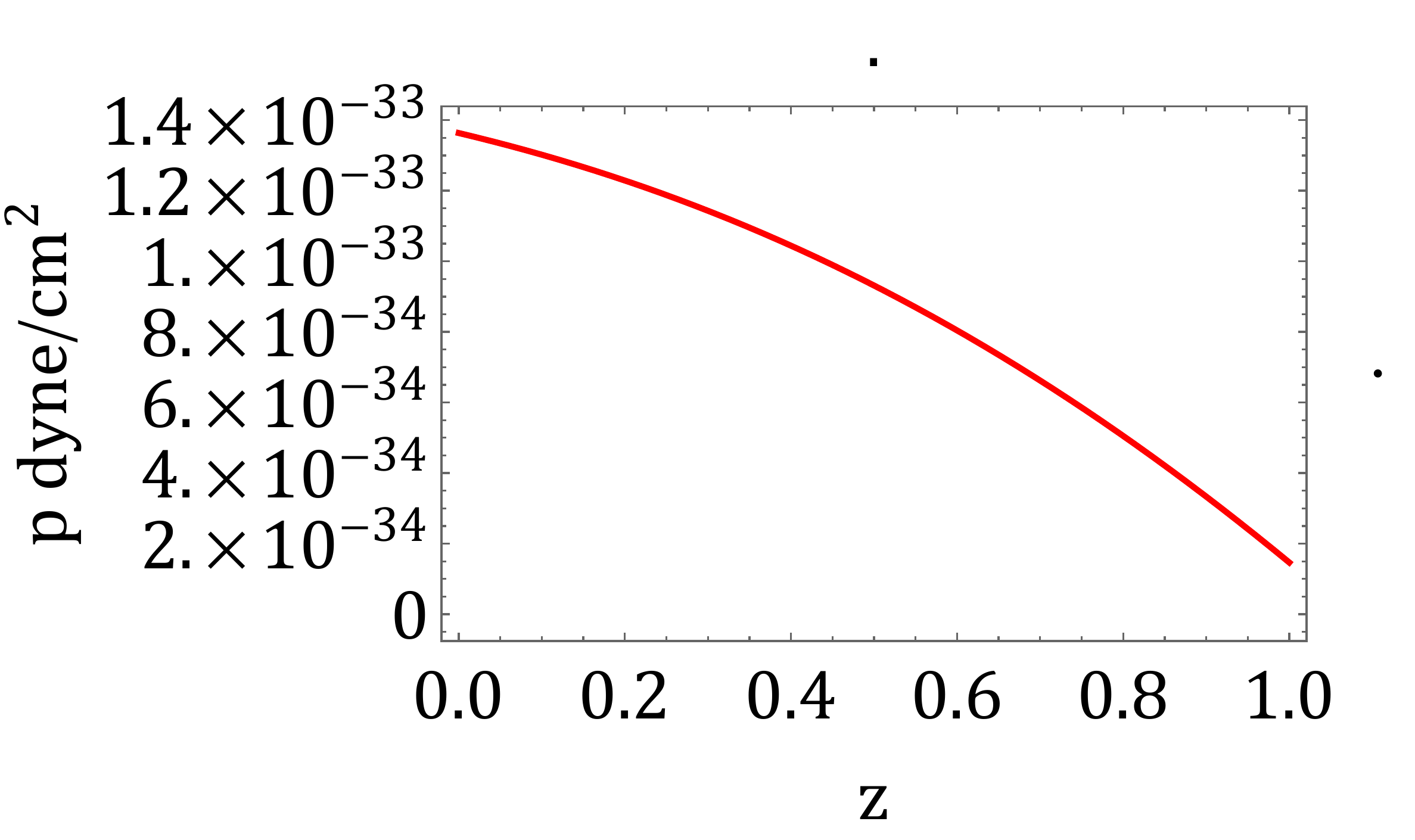}
  	\caption{Figures $(a)$ and $(b)$ are densities plots for our model.  $\mu$ density is higher than baryon density. They show the right path in the sense that in the past the densities were high. They are increasing over redshift which means that they show deceasing trend over time. Figures $(c)$ and $(d)$ are plots for baryon and $\mu$ pressures. baryon pressure is positive where as $\mu$ pressure is negative.  $\rho_{c0}$ is taken  as $1.88\times 10^{-29}h^2 gm/cm^3$, where $h$ = $H_0/100$.}
  	 \end{figure}
   Figures $3(a)$ and $3(b)$ are densities plots for our model.  $\mu$ density is higher than baryon density. They show the right path in the sense that in the past the densities were high. They are increasing over redshift which means that they show deceasing trend over time. Figures $3(c)$ and $3(d)$ are plots for baryon and $\mu$ pressures. baryon pressure is positive where as $\mu$ pressure is negative.  $\rho_{c0}$ is taken as $1.88\times 10^{-29}h^2 gm/cm^3$, where h= $H_0/100$.
   
    \section{Effective Density and Effective Pressure:}
    Equation (\ref{22}) may be written as 
  \begin{equation}\label{47}
  	H^2 (1-2 q)=-8 \pi p_{eff}, ~~ 	3 H^2=8 \pi \rho_{eff} ,
  \end{equation}
  where $ p_{eff} = ( p +  p_{\mu} +p_{\sigma})$ and $\rho_{eff} =( \rho + \rho_{\mu} + \rho_{\sigma}).$ 
  We can  express these in a more convenient way as follows:
  \begin{equation}\label{48}
  	\rho_{eff} = \rho_{c0}	\frac{H^2}{H^2_0}~;~  p_{eff} =  \rho_{c0} 	\frac{H^2(2q-1)}{3H^2_0} .
  \end{equation}
  Figures $4(a)$ is the effective density plot for our model. It shows the right path in the sense that in the past the density was high. It is increasing over redshift which means that it is deceasing over time. Figure $4(b)$ is plot for effective pressure for our model. It  is negative which describe acceleration in the universe.  $\rho_{c0}$ is taken  as $1.88\times 10^{-29}h^2 gm/cm^3$, where h= $H_0/100$.
  
 \begin{figure}[H]
	(a)	\includegraphics[width=9cm,height=8cm,angle=0]{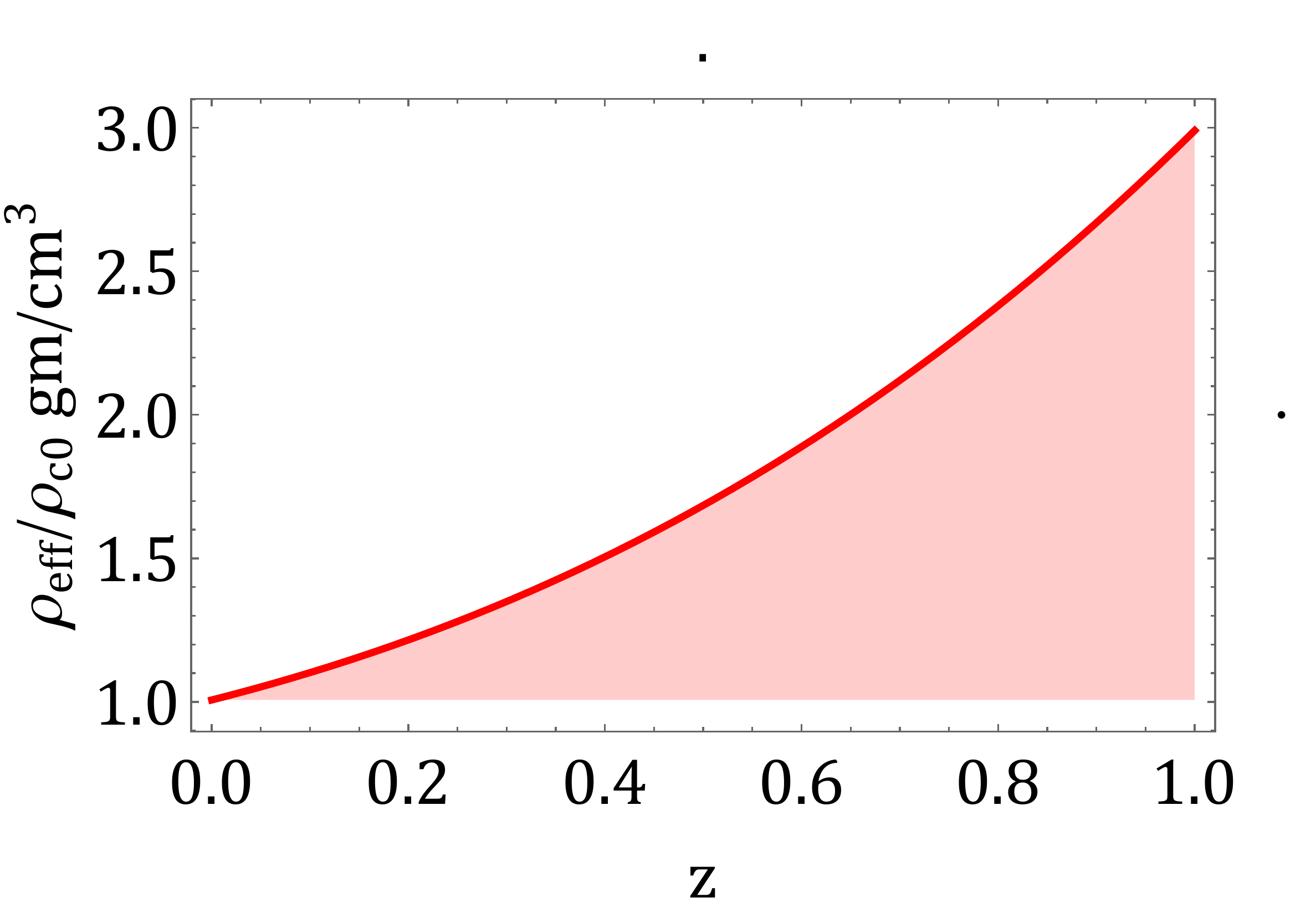}
	(b) \includegraphics[width=9cm,height=8cm,angle=0]{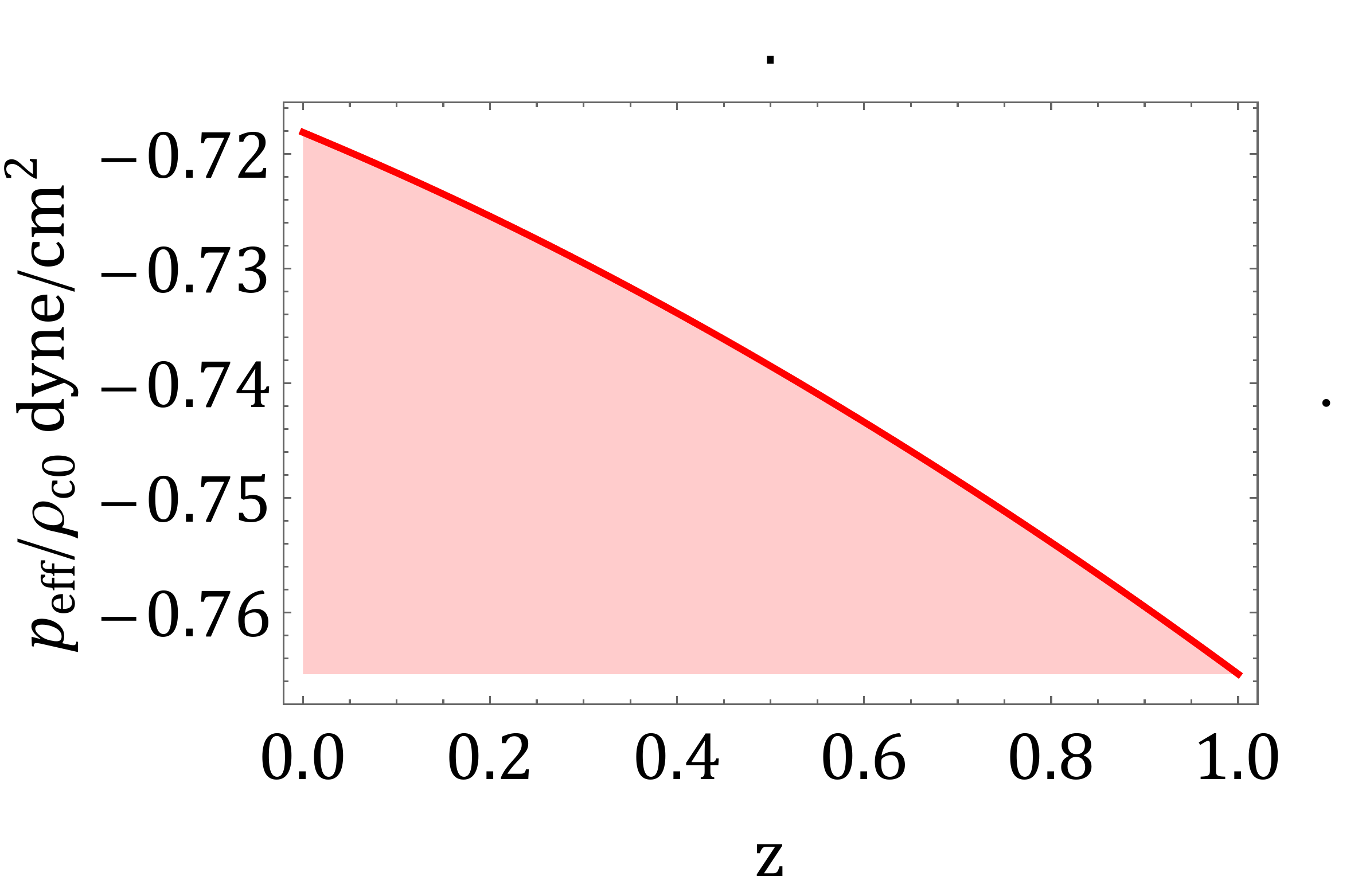}
    \caption{Figure $(a)$ is the effective density plot for our model. It shows the right path in the sense that in the past the density was high. It is increasing over redshift which means that they show deceasing trend over time. Figure $(b)$ is plot for effective pressure for our model. It  is   is negative which describe acceleration in the universe.  $\rho_{c0}$ is taken  as $1.88\times 10^{-29}h^2 gm/cm^3$, where h= $H_0/100$.}
\end{figure}

\subsection{Time versus Red Shift, Age of the Universe and Transitional times:}
We can calculate the time of any event from the red shift through the following transformation
\begin{equation}\label{49}
	(t_0-t_1)=\intop_{t_1}^{t_0}dt=\intop_{a_1}^{a_0}\frac{da}{aH}=\intop_{0}^{z_1}\frac{dz}{(1+z)H(z)},
\end{equation}
where we have used $\frac{a_0}{a}=1+z $ and $ \dot{z}=-(1+z)H.$ $t_0$ and $t_1$ are present and some past time. We note that at present,t=$t_0$ and z=0. 
With the help of expression for Hubble parameter Eq. (\ref{31b}), we observed that 
 there is an asymptote which gives the present age of the universe as $H_0 t_0$= 0.9410. On the basis of the various estimates of $H_0$, the age of the universe comes to $13.5245*10^9$ yrs and  $12.0854 \times 10^9$ yrs in our model. The transition time where the universe inters the accelerating phase is also calculated as $4.10953 \times 10^9$ yrs and  $3.67228 \times 10^9$ yrs as from now.

 \section{Conclusion:} In  a brief nutshell, we say that we have developed a Bianchi I cosmological model of the universe in $f(R,T)$ gravity theory which fit good with the present day scenario of accelerating universe. The main findings of our model are stated point wise.
\begin{itemize}
	\item    The model displays transition from deceleration in the past to the acceleration at the present. 
	\\
	\item  As in the  $\Lambda$CDM model, We have defined the three energy parameters  $\Omega_m$, $\Omega_{\mu}$ and $\Omega_{\sigma}$ such that $\Omega_m$ + $\Omega_{\mu}$ +  $\Omega_{\sigma}$ = 1. The parameter $\Omega_m$ is the matter energy density (baryons + dark matter), $\Omega_{\mu}$ is the energy density associated with the Ricci scalar $R$ and the  trace $T$ of the energy momentum tensor and $\Omega_{\sigma}$ is the energy density associated with the anisotropy of the universe. We shall call $\Omega_{\mu}$  dominant over the other two due to its higher value.  We find that the $\Omega_{\mu}$  and the other two  in the ratio 3:1.\\
	\item   46 Hubble OHD data set is used to estimate present values of Hubble $H_0$, deceleration $q_0$ and jerk $j$ parameters. 1$\sigma$, 2$\sigma$ and 3$\sigma$ contour region plots for the estimated values of parameters  are presented.\\
	\item 580 SNIa supernova distance modulus data set and  66 pantheon SNIa data which include high red shift data in the range $0\leq z\leq 2.36$ have been used to draw error bar plots and likelihood probability curves for distance modulus and apparent magnitude of SNIa supernova's.\\
	 \item  We have  calculated the pressures and densities associated with the two matter densities, viz., $p_{\mu}$, $\rho_{\mu}$, $p_m$ and $\rho_m$, respectively. The present age of the universe as per our model is also evaluated and it is found at par with the present observed values.\\
     \end{itemize}
      The authors are confident that readers and researchers will find our work valuable in the final analysis. The Bianchi I spatially homogeneous and anisotropic model, which fits based on the observational ground, has been constructed within the framework of f(R,T) gravity theory. Our approach in this paper provides a unique and fresh option for seeing the future of the universe.
      
     \section*{ Data Availability Statement:}
     This manuscript has no associated data or the data will not be deposited.
      
      \section*{Acknowledgement}
     The author (A. Pradhan) thanks the IUCAA, Pune, India for providing facilities under associateship programs. The authors acknowledge sincere thanks to anonymous referee for constructive suggestions. 
      
  
 \end{document}